\documentclass[conference]{IEEEtran}
\IEEEoverridecommandlockouts
\usepackage{cite}
\usepackage{amsmath,amssymb,amsfonts}
\usepackage{algorithmic}
\usepackage{graphicx}
\usepackage{textcomp}
\usepackage{xcolor}

\usepackage{hyperref}
\usepackage{arydshln}  % For dashed line in tables
\usepackage{booktabs}
\usepackage{balance}
\usepackage{hhline}
\usepackage{easyReview}
\usepackage{multirow}
\usepackage{textcomp}

\def\BibTeX{{\rm B\kern-.05em{\sc i\kern-.025em b}\kern-.08em
    T\kern-.1667em\lower.7ex\hbox{E}\kern-.125emX}}
    
\begin{document}

\title{Automatic Classification of Magnetic Chirality of Solar Filaments from H-Alpha Observations\\
% \thanks{Identify applicable funding agency here. If none, delete this.}
}

% ------------------------------------------%
%                                           %
%       Azim: Authors with footnotemark     %
%                                           %
% ------------------------------------------%
% \author{
%     \IEEEauthorblockN{Anonymous \IEEEauthorrefmark{1}}
% }

\author{
    \IEEEauthorblockN{
        Alexis~Chalmers\IEEEauthorrefmark{1} and
        Azim~Ahmadzadeh%\IEEEauthorrefmark{1},
    }
    \IEEEauthorblockA{
        Department of Computer Science,
        University of Missouri - St. Louis,
        St. Louis, MO, USA\\
        Email: \IEEEauthorrefmark{1}ajcyrw@umsystem.edu\\
        \vspace{-1.1cm}
    }
}

\maketitle

\begin{abstract}
In this study, we classify the magnetic chirality of solar filaments from H$\alpha$ observations using state-of-the-art image classification models. We establish the first reproducible baseline for solar filament chirality classification on the MAGFiLO dataset. The MAGFiLO dataset contains over $10,000$ manually-annotated filaments from GONG H$\alpha$ observations, making it the largest dataset for filament detection and classification to date. Prior studies relied on much smaller datasets, which limited their generalizability and comparability. We fine-tuned several pre-trained, image classification architectures, including ResNet, WideResNet, ResNeXt, and ConvNeXt, and also applied data augmentation and per-class loss weights to optimize the models. Our best model, ConvNeXtBase, achieves a per-class accuracy of $0.69$ for left chirality filaments and $0.73$ for right chirality filaments. %This study establishes a baseline for solar filament classification, fine-tuned on MAGFiLO, to provide a foundation for future comparable work related to solar filament classification. 

% study provides a foundation for comparfor future work relating to solar filament classification and 

% when using MAGFiLO , which enables comparability due to MAGFiLOs

% and shows how the MAGFiLO dataset enables comparability for future work relating to solar filaments 

\end{abstract}


% ----------------------------------------------------
%   I. INTRODUCTION
% ----------------------------------------------------
\section{Introduction}\label{sec:introduction}
\begin{figure}[t]
        \centering\includegraphics[width=\linewidth]{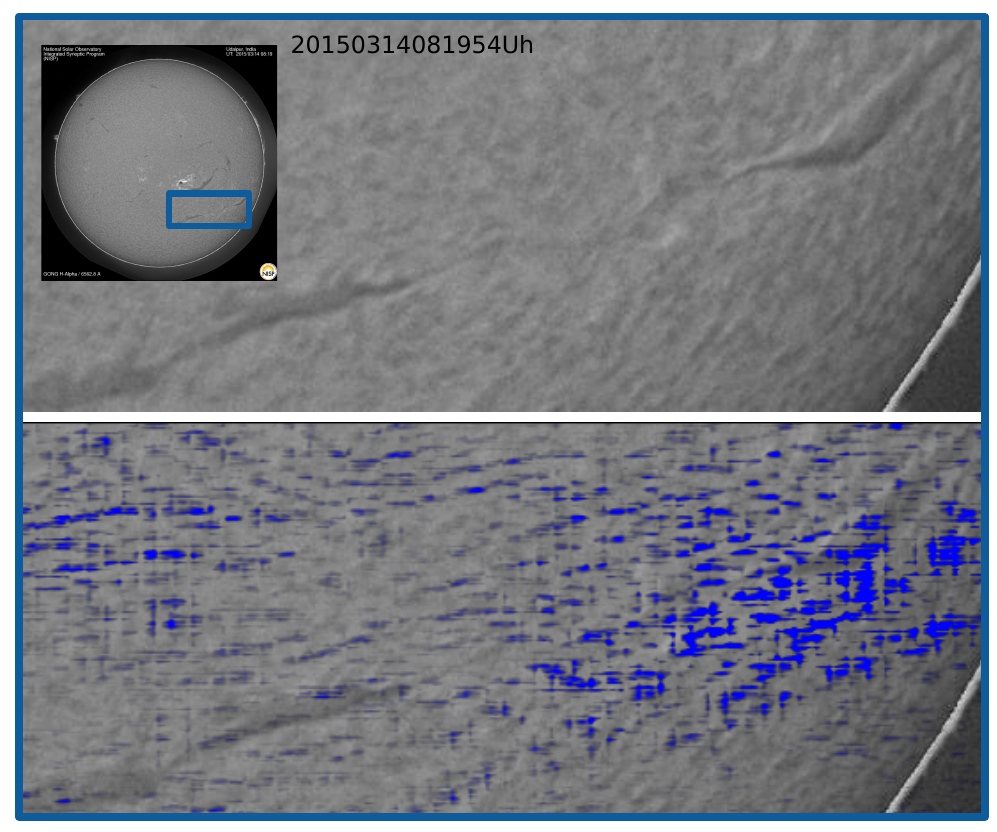}
        \caption{A very large filament captured on March 14, 2015 (08:19:54), is classified by our model as left chiral (sinistral). The overlaid blue map illustrates the pixels of high importance for the classifier.}
        \vspace{-0.3cm}
        \label{fig:cover_page}
    \end{figure}

    Solar filaments are dense structures of plasma suspended in the solar atmosphere by magnetic fields, forming above polarity inversion lines (PILs)—boundaries between positive and negative magnetic regions. When seen on the solar disk in the H$\alpha$ spectral line, they appear dark in absorption (as shown in Fig.~\ref{fig:cover_page}), and when viewed off the limb, they glow in emission as prominences. Filaments often relate to coronal mass ejections (CMEs) and play a key role in solar activity. A subtype called polar crown filaments can indicate polar field reversals, an essential part of the solar magnetic cycle.

    A crucial property of filaments is the chirality (or handedness) of their axial magnetic field. As introduced by Martin \cite{martin1998conditions}, chirality can be inferred from H$\alpha$ images based on filament morphology: if the axial field, viewed from the positive-polarity side of a PIL, points right, the filament is dextral; if left, it's sinistral. Chirality is binary and remains constant through a filament’s lifetime.
        
    Manual identification of solar filaments is a tedious task because there are simply too many of them. Further, solar filaments are fluid, and their 3D structure is not visible in their entirety in H$\alpha$ observations. This makes it very difficult for human experts to confidently identify the chirality of a large portion of filaments. In this study, we explore the effectiveness of the state-of-the-art object classification algorithms to automatically identify the chirality of filaments. To the best of our knowledge, this is one of the few attempts in the exploration of this question on a manually-annotated dataset containing thousands of filaments.
    
% ----------------------------------------------------
%   II. BACKGROUND
% ----------------------------------------------------
\section{Background}\label{sec:background}
    The classification of filaments' chirality has received much less attention (namely, \cite{pevtsov2003chirality, Hazra2018hemispheric, aparna2020solar, kashvi2021analysis, pevtsov2017chirality}) compared with the filament-detection task, i.e., identification of filaments using bounding boxes or pixel-level segmentation (to name a few recent efforts, \cite{guo2022solar, diercke2024universal, liu2021solar, we2022solar}). This could be---at least partially---due to the difficulty of creating a large dataset of manually annotated filaments.

    % TODO: semicolon after "at least one barb were annotated"
    One of the earliest efforts in creating such a dataset was carried out in 2003 \cite{pevtsov2003chirality}. The authors used H$\alpha$ observations (of Big Bear Solar Observatory (BBSO) \cite{zirin1970bbso, denker1999synoptic}) to study the chirality of 2,310 filaments (among other properties) from the year 2000 to 2001. In the rigorous process of creating the referenced dataset, small filaments with no discernible barbs were ignored. All filaments with at least one barb were annotated. However, in the final step, those with only a single barb were excluded regardless of their size.
    
    In 2005, the dataset mentioned above was used to quantify the chirality-classification algorithm \cite{bernasconi2005advanced}. The algorithm, named the Advanced Automatic Filament Detection and Characterization Code (AAFDCC), automatically identified chiralities based on the angle of the identified barbs with respect to the spine of the filament. This was based on the theory introduced earlier in 1998 \cite{martin1998conditions}. The authors of AAFDCC reported a classification accuracy of $0.72$ for predicting filament chirality based on the ground-truth data. To better understand this statistic, two key points should be noted: (1) the ground-truth dataset contained only filaments whose chirality could be discerned by humans, (2) the reported $0.72$ statistic comes from those that were correctly classified (i.e., 473 out of 658); this excludes 170 filaments, which AAFDCC did not assign any label. Additionally, as always, a degree of imperfection must be considered for the ground-truth data, which may result in an unfair underestimation of AAFDCC's accuracy.

    Another example illustrating that the reported classification results should always be taken with a grain of salt is the dataset created in 2018 \cite{Hazra2018hemispheric} where out of the 3,480 filaments examined, the authors managed to identify the chirality of only $765$ filaments ($22\%$). This explains the difficulty of the task. In both \cite{pevtsov2003chirality,Hao2016can} it was also observed that about 20\% of filaments exhibit (at least in still H$\alpha$ images) indications of both left and right chirality. To address this issue, in the MAGFiLO dataset (2024; see Sec.~\ref{subsec:magfilo}), in addition to the Left and Right classes (corresponding to the sinistral and dextral chirality, respectively), a third class label was created, named Unidentifiable, to catch the difficult cases instead of ignoring them. 
    %This should give the reader a better perspective as to how to judge and compare different statistics.

    There have been several other valuable efforts that we do not discuss. In many cases, the algorithms were only evaluated anecdotally on a handful of observations (e.g., \cite{Hao2016can}). In some other studies, the classification task does not concern chirality, e.g., \cite{schuh2014comparative} classifies filaments versus non-filaments, and \cite{guo2022solar} classifies isolated versus non-isolated filaments. We also do not include studies that identify chirality from other data sources such as SDO/AIA, e.g., \cite{Ouyang2017chirality}.

\section{Data}\label{sec:data}
    
    \subsection{Image Data: GONG H-Alpha Observations}\label{subsec:image-gong}
        The Global Oscillation Network Group (GONG) consists of six identical telescopes strategically distributed across the globe: the Big Bear Solar Observatory (California, USA), Mauna Loa Observatory (Hawaii, USA), Learmonth Solar Observatory (Australia), Udaipur Solar Observatory (India), Observatorio del Roque de los Muchachos (Spain), and Cerro Tololo Inter-American Observatory (Chile). This geographic distribution minimizes the effects of Earth’s diurnal cycle on solar observations, enabling near-continuous coverage. The network provides uninterrupted, year-round monitoring of the Sun \cite{hill1994global1, hill1994global2}. In 2021, GONG reported an average duty cycle of 93\%---the proportion of each 24-hour period during which data is collected---a level of continuity that has been consistently sustained over the past 18 years of its 25-year operation \cite{jain2021continuous}.

        An H$\alpha$ filter isolates the narrow spectral line at 656.3 nanometers, produced by hydrogen atoms, one of the most prominent features in the solar spectrum. By transmitting only this wavelength, it enables ground-based observatories to image the solar chromosphere with high contrast. Unlike white-light observations, H$\alpha$ imaging reveals filaments, prominences, and plasma motions shaped by magnetic fields. These observations provide a continuous view of solar activity, supporting both scientific studies of magnetic dynamics and practical applications such as space-weather monitoring.

    \subsection{Gold-standard Annotations: MAGFiLO}\label{subsec:magfilo}
        Efforts involving supervised machine learning have typically encountered the common limitation of insufficient gold-standard data---i.e., datasets of manually-annotated filaments. Due to manual annotation being a time-consuming and costly process, gold-standard data has generally been limited to small datasets (e.g., \cite{zharkova2005filament, shang2024solar}). To overcome this constraint, some researchers have turned to silver-standard data---annotations generated by other filament-detection algorithms---as a basis for training their models (e.g., \cite{zheng2024developing, zhu2019solar,ahmadzadeh2019toward}).  Recently, a large dataset of solar filaments, namely Manually Annotated GONG Filaments from H$\alpha$ Observations (MAGFiLO) \cite{ahmadzadeh2024magfilo, ahmadzadeh2024magfilodata}, addressed this need. MAGFiLO consists of over 10,000 manually annotated filaments from about 1,000 H$\alpha$ observations captured by the GONG network, spanning the years 2011 through 2022. To the best of our knowledge, this is currently the largest manually-annotated dataset of filaments. MAGFiLO comes in the COCO format \cite{lin2014microsoft} and is stored as a JSON file. Each annotation is made up of four pieces: (1) a label representing the filament's magnetic-field chirality (`left', `right', or `unidentifiable'), (2) a polyline representing its spine, (3) a pixel-precise segmentation representing the filament's shape, and (4) a minimal bounding box that tightly inscribes the filament.

    \subsection{Data Preparation}\label{subsec:data-preparation}
        %%- Download FITS images
        %%- Convert FITS to JPEG
        %%- Load JPEG image directory
        %%- Calculate normal distribution
        %%- Normalize images (Dont think this is accurate anymore, need to review code after normalization changes)
        %%- Create core dataset class
        %%- Split dataset into three groups. 
        %%    - Training (70%)
        %%    - Validation (10%)
        %%    - Testing (20%)
        %% - Setup Data loaders 

        % Need for this paper dataset description    
        % Each model compared in this paper is an image classification model with pretrained weights. To fine-tune the models, a dataset of individual filaments was needed. The MAGFiLO dataset is for multi-object detection, classification, and segmentation of solar disk images, but it contains the information needed to generate a dataset for image classification. The dataset needs to be split into training, validation, and testing sets, with each set being balanced between left, right, and unidentifiable chirality classes. The dataset for training image classification models needs to be of images for individual filaments in a directory corresponding to their class. 

        %TODO Wording 
        The MAGFiLO dataset is already prepared for object detection and classification. However, our preliminary study focuses on image classification in which localization is not involved. Therefore, we created a new dataset of images of single filaments. Each filament image is a cropped region from a full-disk H$\alpha$ observation, corresponding to a single filament. The exact region is determined by the corresponding bounding-box data provided in MAGFiLO. We then created three subsets of filament images to be used for training (80\%), validation (10\%), and testing (10\%) the models. We populated these subsets in such a way that the number of instances of the three classes is roughly balanced, mimicking the overall balance in MAGFiLO. The exact number of filaments of each class per set is listed in Table~\ref{tab:filament_counts}.

        % Exact number of images per set and chirality in a table
        \begin{table}[t]
            \centering
            \caption{Counts of filaments of each chirality class in training, validation, and test sets.}
            \begin{tabular}{l|r|r|r||r}
                \hline
                \textbf{Set} & \textbf{Left} & \textbf{Right} & \textbf{Unidentifiable} & \textbf{Total}\\ \hline
                Training      & 2171 (31.75\%)  & 2136 (31.24\%)   & 2530 (37.00\%)     & 6837 \\ \hline
                Validation    & 309 (31.95\%)   & 313 (32.37\%)    & 345 (35.68\%)      & 967 \\ \hline
                Test          & 299 (32.68\%)   & 302 (33.01\%)    & 314 (34.32\%)      & 915 \\ \hline \hline
                \textbf{Total}& 2779 (31.87\%)  & 2751 (31.55\%)   & 3189 (36.57\%)      & 8719 \\ \hline
            \end{tabular}
            \label{tab:filament_counts}
        \end{table}

        %TODO: Look into use of word implicit
        %TODO: Look into paragraph deeper
        Since we only have $6,837$ filaments in our training set, data augmentation was applied to enhance sample diversity. We utilize PyTorch's modules, specifically the DataLoader and ImageFolder modules, to feed images into the classification models in batches \cite{paszke2019pytorch}. In this process, we implicitly use the Transform module to run image transformations, augmenting data. Data augmentation is a common practice for increasing the diversity of data points towards the enhancement of a model's robustness. In Sec.~\ref{classification_with_augmentation_and_class-weights} we show the exact contribution of each of these practices, separately.

        The augmentation process in our pipeline is as follows. Each filament image undergoes controlled, random adjustments of the existing filaments. That is, we do not generate synthetic images, but modify existing images to create seemingly unique images. We apply three transformations, namely, rotation, brightness, and contrast adjustments. The rotation is applied within $\pm 15^{\circ}$ about the center of the filament image, with a probability of 0.5. The brightness and contrast transformations are applied with the adjustment coefficient ranging from 0.6 to 1.2, each with the same probability of 0.5. We do not use transformations that may affect or alter the chirality of a filament. For instance, the flip transformation mirrors a filament and therefore switches the handedness of the filament. Because of the assigned probability and randomness, each image undergoes a combination of these transformations.

\section{Experiments and Results}\label{sec:experiments_and_results}
    We performed multiple experiments to determine the best model architecture, configuration, and data treatment pipeline. Below, we detail our efforts for reproducibility purposes. The experiments were performed on a device with an Intel \textregistered CoreTM i5-10400F CPU @2.90GHz with 32 Gb of memory, and an NVIDIA GeForce RTX 3060 GPU.
    
    \subsection{Models and Metrics}\label{subsec:models}
        To determine the ability of classification models to predict filaments' chirality, a range of different models were tested. Specifically, we tested Resnet (50, 101, 152), WideResnet (50, 101), ResNeXt (50-32, 101-32, 101-64), and ConvNeXt (Base). Pretrained models were fine-tuned to reduce training time. The models all come from PyTorch's image classification models \cite{paszke2019pytorch}. We try four different models with different configurations. 
        
        Of the selected models, three are close derivatives to the ResNet architecture, making ResNet a good baseline. ResNet introduces the use of shortcut connections, which give their output to a layer two or more layers deeper, allowing for a residual feature map to be maintained, solving the problem of vanishing or exploding gradients \cite{he2015deepresiduallearningimage}. Pretrained ResNet models and derivative models are available with depths, number of layers in the model, of 50 and 101, but due to ResNet's focus on thin and deep architectures, it is also available with a depth of 152. WideResNet modifies ResNet by increasing the number of features per convolutional layer while decreasing the depth to maintain the parameter count. Decreasing the depth also decreases the time required to train the model while still outperforming ResNet models \cite{zagoruyko2017wideresidualnetworks}. ResNeXt improves on ResNet by using grouped convolutions where an image is split and each portion of the image is transformed on a different path, then merged back together. The number of paths in a model is its cardinality, which is often $32$ or $64$. ConvNeXt is a recent model, released in 2020, applying the prior design benefits of models like ResNeXt with changes to bring the architecture more in line with transformer models \cite{liu2022convnet2020s}. One major change is that, unlike other models that use Rectified Linear Units (ReLU) as their activation function, ConvNeXt models use Gaussian Error Linear Units (GELU) as their activation function. Similar to the ReLU, GELU has a non-linear output for negative values, but differs in that GELU is smoother than ReLU. This distinction, alongside other modifications, sets the model apart and makes it behave somewhat differently from other models. 

        % \textcolor{red}{Discuss Resnet, ConvNeXt, ResNeXt, and WideResnet, in the order your first list. For models listed in the table, please only mention what `base', `50', `101', `152', `10164' mean.}
        %Resnet model

        To quantify the performance of each model, we report accuracy, precision, recall, and $f_1$ score. All these measures are quantities based on the confusion matrix. The confusion matrix for a binary classification task is a $2\times2$ table with quantities for $tp$ (positive instances correctly classified), $tn$ (negative instances correctly classified), $fp$ (negative instances incorrectly classified), and $fn$ (positive instances incorrectly classified). Accuracy measures the ratio $acc=\frac{tp + tn}{p+n}$ where $p$ and $n$ denote the count of positive and negative instances, respectively. Recall (a.k.a. sensitivity) returns the ratio $rec = \frac{tp}{p}$, and precision returns the ratio $pre=\frac{tp}{tp + fp}$. $f_1$ score (a.k.a. $F$ measure) is the harmonic mean of precision and recall, i.e., $f_1 = 2 \cdot \frac{pre \cdot rec}{pre + rec}$.
    
        Since these measures are designed for binary tasks and our algorithms predict three classes, for multi-class classification problems, it is common to use the one-versus-rest method. For example, once we consider Left as one class and combine Right and Unidentifiable classes into another one. In this setting, true-positive means that a Left filament is correctly classified as Left; true-negative means that a non-Left (Right or Unidentifiable) filament is correctly classified as non-Left; false-positive means that a non-Left filament is classified as Left; and false-negative means that a Left filament is classified as non-Left.

    \subsection{Classification with Augmentation and Class-Weights}\label{classification_with_augmentation_and_class-weights}
        We train all models listed in Sec.~\ref{subsec:models} on the training set discussed in Sec.~\ref{subsec:data-preparation}. We use the default parameters for each model and report the performance of each model tested on the test set. We use the validation set for early stopping and checkpointing. We keep the content of these subsets the same across all different experiments so that our findings are comparable. The performance of each model, without any specific treatment, is reported in Table~\ref{tab:classification_results_per_class}.

        Looking at the results, the untreated models all perform almost the same, except ConvNeXtBase. Excluding the outlying model, the metrics for each remaining model vary from $0.51$ to $0.54$. With a range of $0.02$, changing the underlying architecture between models based closely on ResNet does not seem to make a drastic difference in performance. However, the outlying model, ConvNeXtBase, has the best score for every metric with an accuracy of $0.54$. 
        
        % use of GELU results in better performance. 

        \begin{table}[t]
            \centering
            \caption{Table of classification performance for different models, with augmentation and/or weights. To report performance, the one-versus-rest method is used.  Bold cells indicate the best performance for a treatment.}
            \begin{tabular}{c|c|c|c|c|c}
                \hline
                \textbf{Model} & \textbf{Metric} & \textbf{Base} & \textbf{Augm.} & \textbf{Weights} & \textbf{Both}\\ \hline
                \multirow{4}{6em}{ResNet50} 
                & Acc  & 0.52& 0.55 & 0.52 & 0.56 \\ \cline{2-6}
                & Pre & 0.52& 0.56 & 0.53 & 0.58 \\ \cline{2-6}
                & Rec    & 0.52& 0.55 & 0.52 & 0.56 \\ \cline{2-6}
                & F1        & 0.52 & 0.55 & 0.51 & 0.55 \\ \hline \hline

                \multirow{4}{6em}{ResNet101} 
                & Acc  & 0.52& 0.55 & 0.51 & 0.55 \\ \cline{2-6}
                & Pre & 0.53& 0.55 & 0.52 & 0.58 \\ \cline{2-6}
                & Rec    & 0.53& 0.55 & 0.51 & 0.55 \\ \cline{2-6}
                & F1        & 0.52& 0.55 & 0.51 & 0.53 \\ \hline \hline

                \multirow{4}{6em}{ResNet152} 
                & Acc  & 0.53& 0.55 & 0.53 & 0.54 \\ \cline{2-6}
                & Pre & 0.54& 0.55 & 0.53 & 0.55 \\ \cline{2-6}
                & Rec    & 0.53& 0.55 & 0.53 & 0.54 \\ \cline{2-6}
                & F1        & 0.53& 0.55 & 0.53 & 0.52 \\ \hline \hline

                \multirow{4}{6em}{WideResNet50} 
                & Acc  & 0.51& \textbf{0.57} & 0.47 & 0.53 \\ \cline{2-6}
                & Pre & 0.51& \textbf{0.58} & 0.48 & 0.59 \\ \cline{2-6}
                & Rec    & 0.51& \textbf{0.57} & 0.47 & 0.53 \\ \cline{2-6}
                & F1        & 0.51& \textbf{0.57} & 0.47 & 0.51 \\ \hline \hline

                \multirow{4}{6em}{WideResNet101} 
                & Acc  & 0.52& 0.56 & 0.48 & 0.52 \\ \cline{2-6}
                & Pre & 0.53& 0.56 & 0.49 & 0.56 \\ \cline{2-6}
                & Rec    & 0.52& 0.56 & 0.48 & 0.52 \\ \cline{2-6}
                & F1        & 0.52& 0.56 & 0.48 & 0.50 \\ \hline \hline

                \multirow{4}{6em}{ResNext50-32} 
                & Acc  & 0.53& 0.56 & 0.50 & \textbf{0.56} \\ \cline{2-6}
                & Pre & 0.53& 0.56 & 0.51 &\textbf{0.56} \\ \cline{2-6}
                & Rec    & 0.53& 0.56 & 0.50 & \textbf{0.56} \\ \cline{2-6}
                & F1        & 0.53& 0.56 & 0.50 & \textbf{0.56} \\ \hline \hline

                \multirow{4}{6em}{ResNext101-32} 
                & Acc  & 0.53& 0.55 & \textbf{0.53} & 0.53 \\ \cline{2-6}
                & Pre & 0.53& 0.56 & \textbf{0.55} & 0.55 \\ \cline{2-6}
                & Rec    & 0.53& 0.55 & \textbf{0.53} & 0.53 \\ \cline{2-6}
                & F1        & 0.53& 0.55 & \textbf{0.53} & 0.51 \\ \hline \hline

                \multirow{4}{6em}{ResNext101-64} 
                & Acc  & 0.53& 0.56 & 0.52 & 0.55 \\ \cline{2-6}
                & Pre & 0.53& 0.56 & 0.52 & 0.56 \\ \cline{2-6}
                & Rec    & 0.53& 0.56 & 0.52 & 0.55 \\ \cline{2-6}
                & F1        & 0.52& 0.56 & 0.52 & 0.54 \\ \hline \hline

                \multirow{4}{6em}{ConvNeXtBase} 
                & Acc  & \textbf{0.54}& 0.44 & 0.46 & 0.50 \\ \cline{2-6}
                & Pre & \textbf{0.55}& 0.42 & 0.53 & 0.59 \\ \cline{2-6}
                & Rec    & \textbf{0.54}& 0.44 & 0.46 & 0.50 \\ \cline{2-6}
                & F1        & \textbf{0.54}& 0.41 & 0.42 & 0.43 \\ \hline \hline

            \end{tabular}
            \label{tab:classification_results_all_class}
        \end{table}

        % Impact of Augmentation on Performance
        Since we observed that models overfit very quickly---after only a few epochs---we added two treatments, namely, augmentation and using classification weights. Overfitting indicates that the model is failing to generalize and is instead memorizing the training data, under-performing on the validation data. To hinder the model from memorizing the data, augmentations were applied (only) to the training images. As shown in Table~\ref{tab:classification_results_all_class}, it is evident that the augmentations improved the generalization capabilities of every model except for ConvNeXtBase (the previously best model with no treatment). Augmentation raised the average classification performance of all models from F1 of $0.52$ to $0.54$.

        % After looking over the macro average values shown in Table~\ref{tab:classification_results_all_class}, the accuracy of models trained on unaugmented data and augmented data did improve but did not reflect the improvement seen in corresponding stacked bar plots. 
        
        % Augmentations must not change the class of the filament. Filament chirality is sensitive to many common transformations. For example, flipping a filament either vertically or horizontally can change the chirality. This quality of filaments limits which transformations can be applied. The brightness and contrast of a filament image are varied slightly, and small rotations do not alter the resulting chirality. 
        
        % Impact of Class-Wight on Performance
        As we mentioned earlier, in the MAGFiLO dataset, the third class label, i.e. Unidentifiable, serves as a bucket for difficult or impossible cases: when an annotator sees contradictory patterns (indicating both left and right chirality) or when there are not enough textural details on/around a filament to discern its chirality, they assigned the Unidentifiable label to such filaments. We take advantage of this design and use classification weights (per class) to handle difficult cases. Assigning a lower weight to the Unidentifiable class decreases the learning penalty for misclassification of those filaments. Conversely, the penalty for the left-chiral filaments identified as right, or vice versa, increases, pushing the model to focus on the ``identifiable'' filaments.

        To investigate this, we assigned the neutral weight coefficient of 1.0 to both the Left and Right classes, and the weight coefficient of $0.25$ to the Unidentifiable class, for all models. Weight coefficients work by multiplying the per-class loss by different values to apply a controlled bias to the model during training. The coefficients $1.0, 1.0, 0.25$ for Left, Right, and Unidentifiable filaments, reduce the misclassification of the Unidentifiable filaments down to one quarter of the penalty of misclassification for the other two classes. As shown in Table~\ref{tab:classification_results_all_class}, this treatment alone is not as impactful as the augmentation treatment, however, using weights combined with augmentation made ResNext50-32 compete with WideResNet50, with F1 scores of $0.57$ and $0.56$, respectively.

        Because of the intrinsic difficulty in classifying the chirality of Unidentifiable filaments, the quantities shown in Table~\ref{tab:classification_results_all_class} may be obscuring the reality, due to the averaging effect. Looking at per-class accuracy, as shown in Table~\ref{tab:classification_results_per_class}, reveals the true performance of these models: with augmentation and weights (1.0 for Left, 1.0 for Right, and 0.25 for Unidentifiable), the filaments that are successfully annotated by trained humans are annotated correctly 68.65\% (Left) and 72.85\% (Right) of the time. Under such a configuration, the model is only 8\% correct about the Unidentifiable filaments. That is, almost all such filaments are classified as Left or Right. This is, of course, by design since we assigned the weight coefficient of 0.25 to this class. But it should be noted that these filaments were already flagged, because their chirality was not visible from H$\alpha$ images.

    \subsection{Finding Best Weight Coefficient}
        To investigate the best weight coefficients, we took the best model according to per-class performance (as shown in Table~\ref{tab:classification_results_per_class}), i.e., ConvNeXtBase, and applied a range of weight coefficients $\{0.0, 0.1, 0.2, 0.25, 0.3, 0.4, 0.5\}$. Weights below $0.2$ resulted in the model never predicting the Unidentifiable class. When the weight was set to $0.1$, the model had slightly higher accuracy for right chirality filaments than left chirality filaments, whereas when the weight was set to $0.0$ the model predicted every filament's chirality to be left. The best performing weight was $0.25$ with an accuracy of $0.69$ followed by a weight of $0.2$ with an accuracy of $0.66$. This indicates that the optimal weight is likely around that range. However, there was a trend as weight increased from $0.3$ to $0.5$, accuracy increased from $0.47$ to $0.63$, indicating a potentially different optimal weight. Another trend is that as the weight increases from $0.2$ to $0.5$, the precision of the unidentifiable filaments decreased from $0.87$ to $0.65$. 
        
        \begin{table}[t]
            \centering
            \caption{Table of per-class classification performance for different models, with augmentation and/or weights.  Bold cells indicate an overall better performance.}
            \begin{tabular}{c|c|c|c|c|c}
                \hline
                \textbf{Model} & \textbf{Class} & \textbf{Base} & \textbf{Augm.} & \textbf{Weights} & \textbf{Both}\\ \hline
                \multirow{4}{6em}{ResNet50} 
                & L  & 0.44 & 0.51 & 0.41 & 0.69 \\ \cline{2-6}
                & R  & 0.52  & 0.54 & 0.47 & 0.66 \\ \cline{2-6}
                & U  & 0.60 & 0.60 & 0.67 & 0.33 \\ \hline \hline

                \multirow{4}{6em}{ResNet101} 
                & L  & 0.55 & 0.59 & 0.44 & 0.69 \\ \cline{2-6}
                & R  & 0.45 & 0.49 & 0.45 & 0.67 \\ \cline{2-6}
                & U  & 0.57 & 0.57 & 0.64 & 0.30 \\ \hline \hline

                \multirow{4}{6em}{ResNet152} 
                & L  & 0.53& \textbf{0.61} & 0.53 & 0.63 \\ \cline{2-6}
                & R  & 0.45& \textbf{0.50} & 0.50 & 0.64 \\ \cline{2-6}
                & U  & 0.62& \textbf{0.54} & 0.56 & 0.33 \\ \hline \hline

                \multirow{4}{6em}{WideResNet50} 
                & L  & 0.53& 0.60 & 0.48 & 0.71 \\ \cline{2-6}
                & R  & 0.49& 0.48 & 0.37 & 0.66 \\ \cline{2-6}
                & U  & 0.53& 0.64 & 0.55 & 0.23 \\ \hline \hline

                \multirow{4}{6em}{WideResNet101} 
                & L  & 0.45& 0.53 & 0.37 & 0.70 \\ \cline{2-6}
                & R  & 0.45& 0.50 & 0.42 & 0.62 \\ \cline{2-6}
                & U  & 0.67& 0.64 & 0.65 & 0.24 \\ \hline \hline

               \multirow{4}{6em}{ResNext50-32} 
                & L  & 0.49& 0.59 & 0.43 & 0.65 \\ \cline{2-6}
                & R  & 0.52& 0.54 & 0.45 & 0.56 \\ \cline{2-6}
                & U  & 0.57& 0.55 & 0.62 & 0.48 \\ \hline \hline

                \multirow{4}{6em}{ResNext101-32} 
                & L  & 0.49 & 0.57 & 0.46 & 0.61 \\ \cline{2-6}
                & R  & 0.46& 0.47 & 0.49 & 0.65 \\ \cline{2-6}
                & U  & 0.64& 0.62 & 0.66 & 0.32 \\ \hline \hline
 
                \multirow{4}{6em}{ResNext101-64} 
                & L  & 0.51& 0.57 & 0.52 & 0.66 \\ \cline{2-6}
                & R  & 0.46& 0.50 & 0.47 & 0.62 \\ \cline{2-6}
                & U  & 0.61& 0.61 & 0.57 & 0.36 \\ \hline \hline

                \multirow{4}{6em}{ConvNeXtBase} 
                & L  & \textbf{0.54} & 0.14 & \textbf{0.60} & \textbf{0.69} \\ \cline{2-6}
                & R  & \textbf{0.46} & 0.51 & \textbf{0.64} & \textbf{0.73} \\ \cline{2-6}
                & U  & \textbf{0.63}& 0.67 & \textbf{0.14} & \textbf{0.08} \\ \hline \hline
                
            \end{tabular}
            \label{tab:classification_results_per_class}
        \end{table}

        \begin{table}[t]
            \centering
            \caption{Table of classification performance for ConvNeXtBase (best model) for different weight coefficients. The one-versus-rest method is used.  Bold cells indicate an overall better performance.}
            \begin{tabular}{c|c|c|c|c|c}
                \hline
                \textbf{Weight} & \textbf{Class} & \textbf{Accuracy} & \textbf{Precision} & \textbf{Recall} & \textbf{F1}\\ \hline
                \multirow{4}{2em}{0.0} 
                & L  & 1.00 & 0.33 & 1.00 & 0.49 \\ \cline{2-6}
                & R  & 0.00 & 0.00 & 0.00 & 0.00 \\ \cline{2-6}
                & U  & 0.00 & 0.00 & 0.00 & 0.00 \\ \hline \hline

                \multirow{4}{2em}{0.1} 
                & L  & 0.44 & 0.24 & 0.44 & 0.31 \\ \cline{2-6}
                & R  & 0.50 & 0.40 & 0.50 & 0.44 \\ \cline{2-6}
                & U  & 0.00 & 0.00 & 0.00 & 0.00 \\ \hline \hline

                \multirow{4}{2em}{0.2} 
                & L  & 0.66 & 0.42 & 0.66 & 0.51 \\ \cline{2-6}
                & R  & 0.64 & 0.46 & 0.64 & 0.54 \\ \cline{2-6}
                & U  & 0.06 & 0.87 & 0.06 & 0.12 \\ \hline \hline
  
                \multirow{4}{2em}{0.25} 
                & L  & \textbf{0.69} & 0.46 & \textbf{0.69} & \textbf{0.55} \\ \cline{2-6}
                & R  & \textbf{0.73} & 0.50 & \textbf{0.73} & \textbf{0.59} \\ \cline{2-6}
                & U  & \textbf{0.08} & 0.81 & \textbf{0.08} & \textbf{0.15} \\ \hline \hline

                \multirow{4}{2em}{0.3} 
                & L  & 0.47 & 0.28 & 0.47 & 0.35 \\ \cline{2-6}
                & R  & 0.47 & 0.42 & 0.47 & 0.45 \\ \cline{2-6}
                & U  & 0.16 & 0.77 & 0.16 & 0.26 \\ \hline \hline

               \multirow{4}{2em}{0.4} 
                & L  & 0.59 & 0.43 & 0.59 & 0.50 \\ \cline{2-6}
                & R  & 0.58 & 0.45 & 0.58 & 0.51 \\ \cline{2-6}
                & U  & 0.26 & 0.70 & 0.26 & 0.38 \\ \hline \hline

                \multirow{4}{2em}{0.5} 
                & L  & 0.63 & \textbf{0.49} & 0.63 & 0.55 \\ \cline{2-6}
                & R  & 0.64 & \textbf{0.51} & 0.64 & 0.57 \\ \cline{2-6}
                & U  & 0.32 & \textbf{0.65} & 0.32 & 0.43 \\ \hline \hline
            \end{tabular}
            \label{tab:classification_results_convnext_per_class_by_weight}
        \end{table}

        \begin{figure}[t]
            \centering
            \includegraphics[width=0.9\linewidth]{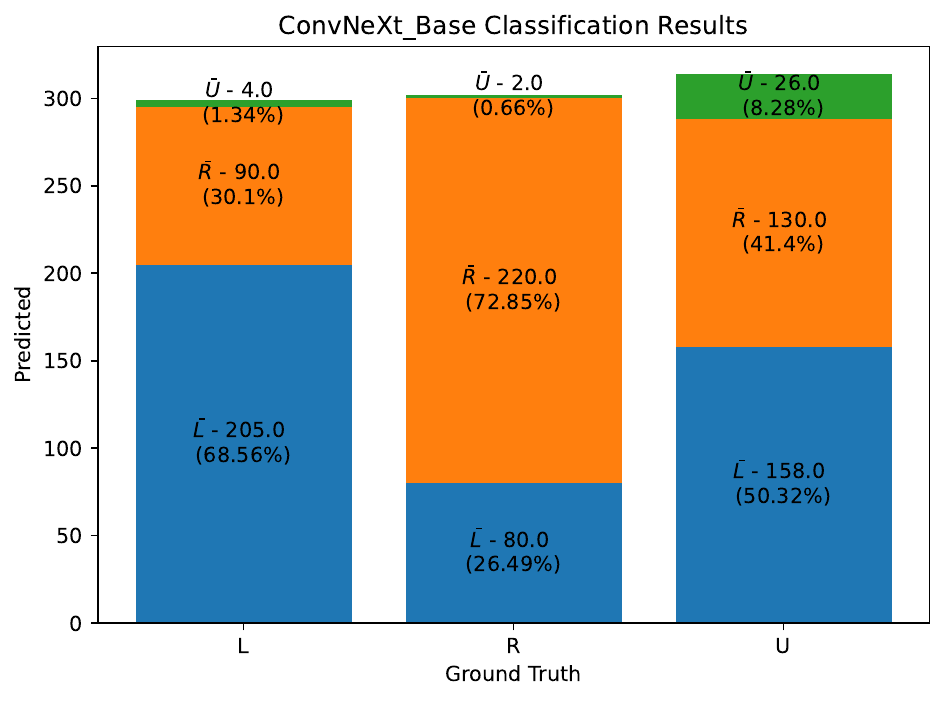}
            \caption{Stacked bar plot showing the ground truth chirality on the $x$ axis and the number of predicted filaments of each chirality for that ground truth on the $y$ axis. Each bar has three segments, one for each predictable chirality. On each segment of a bar, a letter with a hat indicates the predicted chirality, the integer number shows the number of times that chirality was predicted, and the percentage shows the portion of ground truth filaments that were predicted as that chirality.}
            \label{fig:stackedbarplot}
        \end{figure}

        Looking at the best performance, ConvNeXtBase achieves a per-class classification performance of 0.69 for left chirality filaments and 0.73 for right chirality filaments. This is very promising, indicating that the classification model is capable of discerning meaningful textural information about filaments from H$\alpha$ images, without being guided towards filaments' spines and barbs. Furthermore, the balance between precision and recall implies a balance between type I error and type II error. This is desired because of the trade-off between type I and type II errors.

        To understand the true performance of this model, one must take the confidence of human annotators when creating the dataset into account. After all, any model is bound by the quality of the data. As discussed in \cite{ahmadzadeh2024magfilo}, the annotators did not fully agree on the labels they assigned to the same filaments. For model training, this means the model observes contradictory patterns. Considering all three labels (Left, Right, Unidentifiable), the annotators achieved a kappa score of 0.47 (as reported in the MAGFiLO paper). This roughly corresponds to a 65\% match between three labels. To put this in context, in 100 trials of randomly choosing 2 out of three labels, the probability of a match in more than 50\% of the time is almost zero, i.e., $1-F(50; 100, 0.33) \approx 0.0003$. Excluding the Unidentifiable label, the authors claimed that this raised the kappa score to 0.66, which corresponds to an 83\% match between the two labels. Again, but in contrast, the probability of randomly seeing more than 50 matches in 100 binary trials is $1-F(50; 100, 0.5) \approx 0.46$.
        
        To compare the model with humans, we changed the definition of correct classification to be agreeing with \textit{at least} one of the three annotators. For example, if the model predicts a filament's chirality as Left, and at least one of the annotators assigned this label to that filament, we count that as a correct classification. The reasoning behind this experiment is that if agreeing with at least one annotator is counted as a correct classification, we only penalize the model for disagreeing with human annotators, whether annotators agree with each other or not. Thus, the true penalty applies only when the model disagrees with humans, beyond the potential disagreement between humans. The model had a $0.76$ agreement ratio with human annotators. 

% ----------------------------------------------------
%   IV. Results and Evaluation
% ----------------------------------------------------
\section{Qualitative Evaluation and Discussion}\label{sec:qual_results}
    Unlike computer vision techniques, which rely on extracting handcrafted features from images, image classifiers are trained on images and determine feature importance through iterative adjustments. This means that the reason why a model makes a given prediction is unknown without further investigation. Interpretability techniques are used to provide information on why models make the predictions they do. However, most interpretability techniques do not explicitly determine the reason a model makes any given prediction. Instead, they provide information such as heatmaps, which are easier for humans to analyze, but are prone to bias since they show where in an image is important but not why. One method to generate a heatmap for interpretability is called Integrated Gradients \cite{sundararajan2017axiomaticattributiondeepnetworks}, which attributes a model's prediction to input features, indicating the importance of each feature in an input. 

\begin{figure*}
        \centering
        \includegraphics[width=1\linewidth]{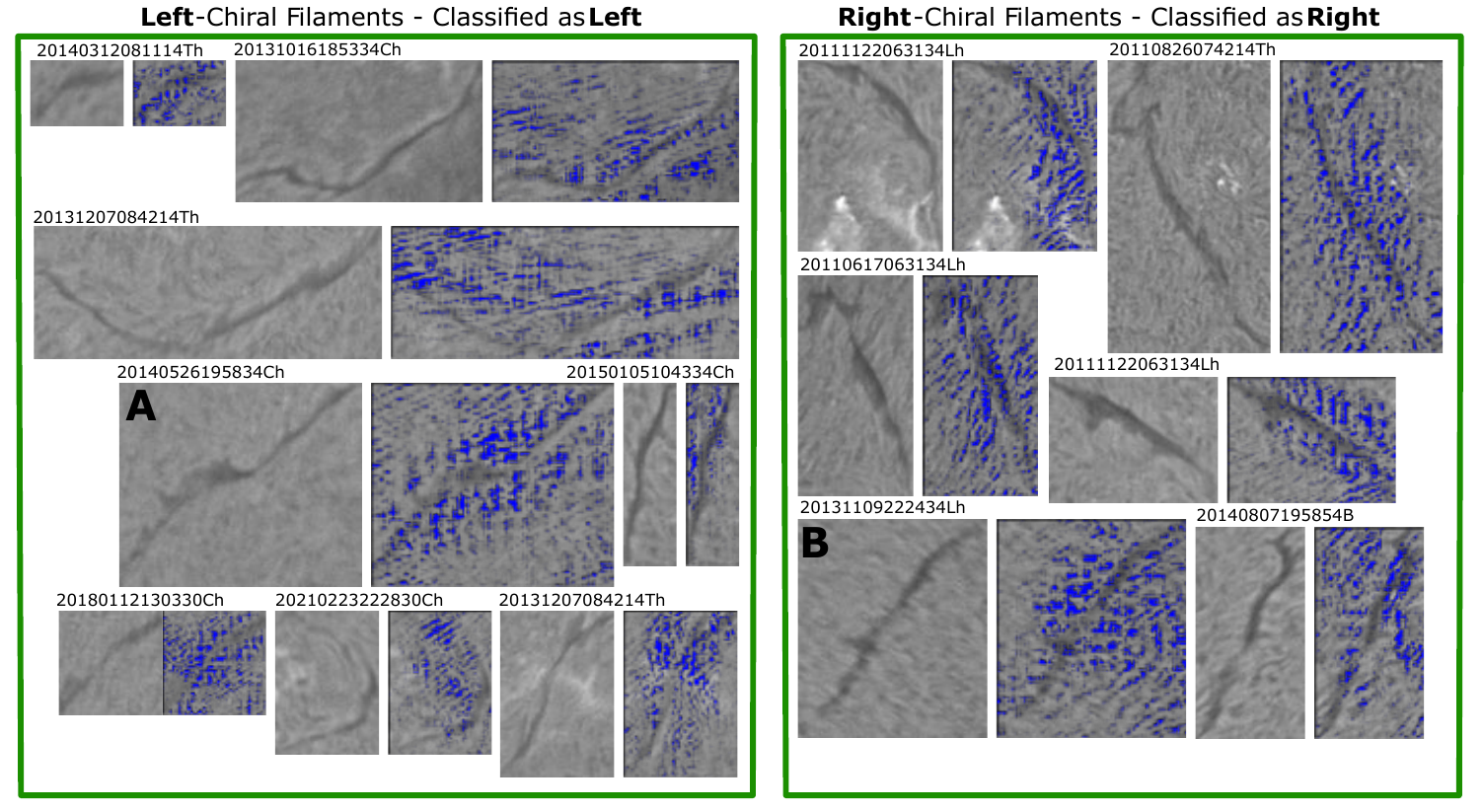}
        \includegraphics[width=1\linewidth]{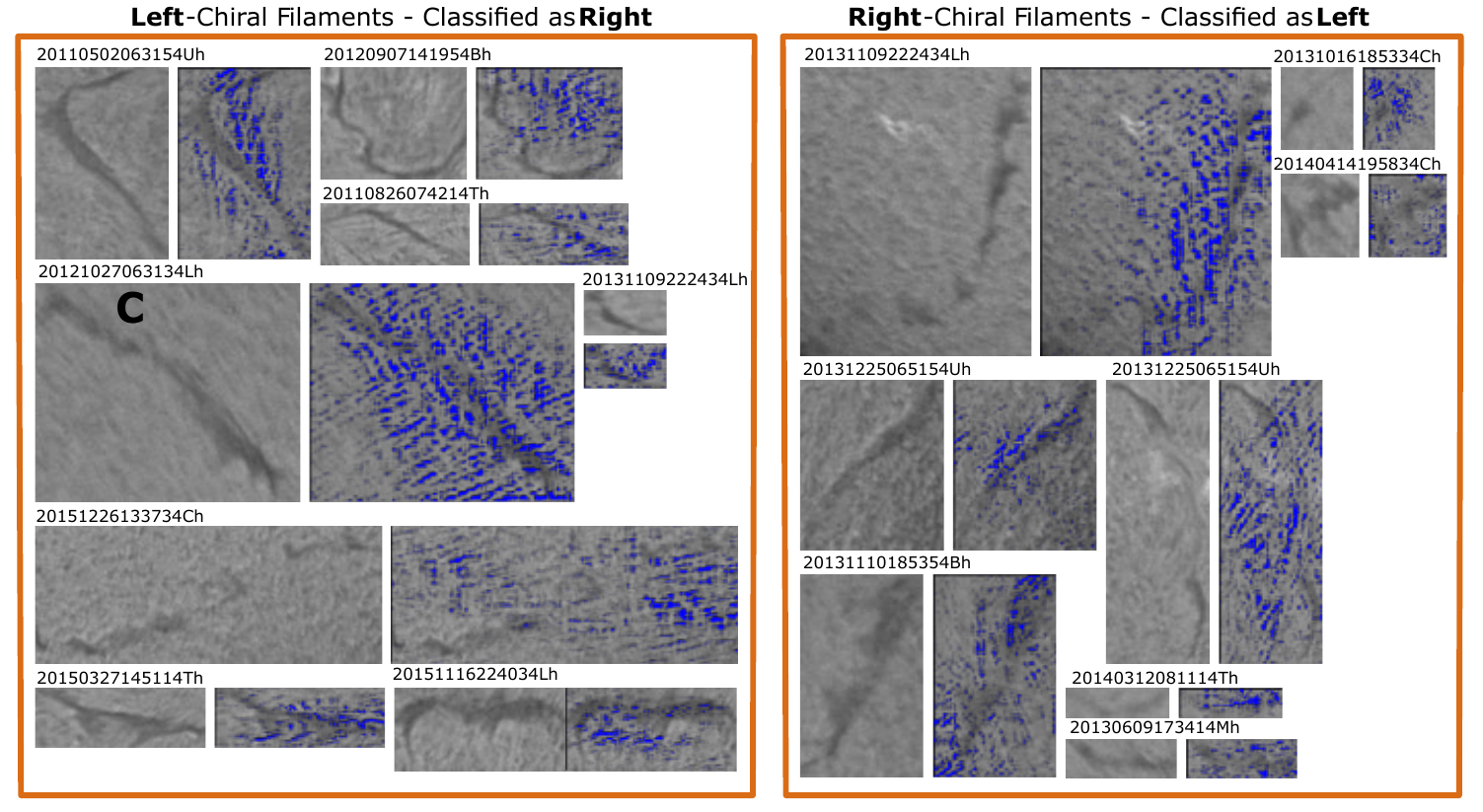}
        \caption{Probability maps of a few filaments classified correctly (top) or incorrectly (bottom) are illustrated. These correspond to classifications made by ConvNeXtBase and are generated using integrated gradients.}
        \label{fig:example_prob_maps}
    \end{figure*}

    Heatmaps of correctly classified filaments generally indicate low importance for regions of the image showing the filament, like filament A in Figure~\ref{fig:example_prob_maps}, showing low importance where the filament plateaus, and following when the filament is diagonal. The filament heatmap indicates importance on the edge of the filament and the surrounding area. This could be due to the model using barbs to determine a filament's chirality. However, correctly classified filaments can have heatmaps indicating importance in regions not near the actual filament, like filament B in ~\ref{fig:example_prob_maps}, where the bottom right region does not contain any filaments, but the heatmap shows high importance. Incorrectly classified filaments have heatmaps indicating importance in the imaged filament's area or far away from the filament, but not at the filament's edge. Filament C in Figure~\ref{fig:example_prob_maps} indicates importance for the filament's area and a wide area surrounding the filament, but has a narrow band, which is indicated as not important at the filament's edge. The primary difference in heatmaps between a correctly and incorrectly classified filament is whether or not the edge is important, meaning the model likely uses the edge to determine chirality.

% ----------------------------------------------------
%   V. Conclusion and Future Work
% ----------------------------------------------------
\section{Conclusion and Future Work}\label{sec:conclusion}

    Automatic classification of filament chirality is a question where, due to the difficulty in creating a large manually annotated dataset of filament chiralities, the results of many studies lack a standard dataset that allows for comparison between them. Pretrained image classification models were fine-tuned on cropped regions of full-disk H$\alpha$ observations, which contained filaments, to establish a baseline of performance for comparison. Augmentation and per-class loss weight were used to optimize models by hindering overfitting and penalizing the model less for misclassifying unidentifiable filaments, respectively.

\section*{Acknowledgment}
    This material is based upon work supported by the National Science Foundation under Grant No. 2209912 and 2433781, directorate for Computer and Information Science and Engineering (CSE), and Office of Advanced Cyberinfrastructure (OAC), and Grant No. 2511630, AST Division Of Astronomical Sciences and MPS Directorate for Mathematical and Physical Sciences.
    
    This work utilizes GONG data obtained by the NSO Integrated Synoptic Program, managed by the National Solar Observatory, which is operated by the Association of Universities for Research in Astronomy (AURA), Inc. under a cooperative agreement with the National Science Foundation and with contribution from the National Oceanic and Atmospheric Administration. The GONG network of instruments is hosted by the Big Bear Solar Observatory, High Altitude Observatory, Learmonth Solar Observatory, Udaipur Solar Observatory, Instituto de Astrofísica de Canarias, and Cerro Tololo Interamerican Observatory.

\balance
\bibliographystyle{IEEEtran}
\bibliography{main}

\end{document}